\documentclass{desyproc}

\usepackage{amsmath,amsfonts}
\usepackage{import}
\usepackage{listings}

\usepackage{url}
\usepackage{tabularx}
\usepackage{booktabs}
\usepackage{slashed}
\def\SymbReg{\raisebox{0.4em}{\scalebox{0.5}{\textsuperscript{\bf\textregistered}}}{ }}
\def\SymbTra{\raisebox{0.3em}{\scalebox{0.5}{\textsuperscript{\bf\texttrademark}}}{ }}

\begin{document}
\title{Conjugate gradient solvers on Intel\SymbReg Xeon Phi\SymbTra and NVIDIA\SymbReg GPUs}

\author{{\slshape O.\ Kaczmarek$^1$, C.\ Schmidt$^1$, P.\ Steinbrecher$^1$ and M.\ Wagner$^2$}\\[1ex]
$^1$Fakult\"at f\"ur Physik, Universit\"at Bielefeld, D-33615 Bielefeld, Germany\\
$^2$Physics Department, Indiana University, Bloomington, IN 47405, USA}

\contribID{11}

\confID{7534}  
\desyproc{DESY-PROC-2014-05}
\acronym{GPUHEP2014} 
\doi  

\maketitle

\begin{abstract}
Lattice Quantum Chromodynamics simulations typically spend most of the runtime in inversions of the Fermion Matrix. 
This part is therefore frequently optimized for various HPC architectures. Here we compare the performance of the Intel\SymbReg Xeon Phi\SymbTra to current Kepler-based NVIDIA\SymbReg Tesla\SymbTra GPUs running a conjugate gradient solver. By exposing more parallelism to the accelerator through inverting multiple vectors at the same time, we obtain a performance greater than $300\;\mathrm{GFlop/s}$ on both architectures. This more than doubles the performance of the inversions. We also give a short overview of the Knights Corner architecture, discuss some details of the implementation and the effort required to obtain the achieved performance.
\end{abstract}

\section{Introduction}
In finite temperature Quantum Chromodynamics (QCD) fluctuations of conserved charges, baryon number ($B$), electric charge ($Q$) and strangeness ($S$), are particular interesting observables. They can be measured in experiments at the Relativistic Heavy Ion Collider (RHIC) and the Large Hadron Collider (LHC) and have also been calculated in Lattice QCD (LQCD) with increasing precision~\cite{Bazavov:2014yba_xya}. They are derived 
from generalized susceptibilities
\begin{align}
    \chi^{BQS}_{mnk}(T) =  \left.\frac{1}{VT^3}  
    \frac {\partial^{m+n+k} \ln \mathcal{Z}} {
    \partial \left( \mu_B / T \right)^m
    \partial \left( \mu_Q / T \right)^n
    \partial \left( \mu_S / T \right)^k
    }
    \right|_{\vec{\mu}=0}\ ,
\end{align}
where $\mathcal{Z}$ denotes the partition function of the medium at temperature $T$ and volume $V$. 

In LQCD the required derivatives of $\mathcal{Z}$ w.r.t.\ the chemical potentials $\mu$ can be obtained by stochastically estimating traces over combinations of the inverse and derivatives of the Fermion Matrix $M$ with a sufficiently large number of random vectors $\eta$, e.g.
\begin{align}
\operatorname{Tr} \left(\frac{\partial^{n_1} M}{\partial \mu^{n_1}} M^{-1} \frac{\partial^{n_2} M}{\partial \mu^{n_2}} \ldots M^{-1}\right) = \lim_{N \rightarrow \infty} \frac{1}{N} \sum_{k=1}^{N} \eta_k^\dagger \frac{\partial^{n_1} M}{\partial \mu^{n_1}} M^{-1} \frac{\partial^{n_2} M}{\partial \mu^{n_2}} \ldots M^{-1} \eta_k\ .
\end{align}
To control the errors we use 500-1500 random vectors on each gauge configuration. Depending on the desired highest derivative degree this involves several inversion of the Fermion Matrix for each random vector.

For reasons of the numerical costs, staggered fermions are the most common type of fermions for thermodynamic calculations on the lattice. We use the highly improved staggered fermion (HISQ) action~\cite{Follana:2004HISQ}. In terms of the smeared links $U$ and Naik links $N$ the Dslash operator reads
\begin{align}
    w_x = D_{x,x'} v_{x'} 
        = \sum_{\mu=0}^4
          \left[ \left (U_{x,\mu} v_{x+\mu}- U^\dagger_{x-\mu,\mu } v_{x-\mu}\right) + 
          \left(N_{x,\mu} v_{x+3\mu}- N^\dagger_{x-3\mu,\mu } v_{x-3\mu}\right)
          \right]\ .
\end{align}
Here $N$ and $U$ are complex $3\!\!\times\!\!3$ matrices and $v,w$ are complex $3$-dimensional vectors.
Within the inversion the application of the Dslash operator typically consumes more than $80\%$ of the runtime. This part already has a low arithmetic intensity (see Tab.~\ref{arithm_intens}) and the average arithmetic intensity is further decreased by the linear algebra operation in the conjugate gradient. Thus, the achievable performance is clearly bound by the available memory bandwidth. 
Given its massively parallel nature and the bandwidth hunger it is well suitable for accelerators.
Lattice QCD simulations make extensive use of GPUs for several years now~\cite{gpu}. The MIC architecture is also gaining more attraction and codes are being optimized~\cite{phi}.
A common optimization to reduce memory accesses in LQCD is to exploit available symmetries and reconstruct the gauge links from 8 or 12 floats instead of loading all 18 floats. For improved actions these symmetries are often broken and thus we can only reconstruct the Naik links from $9$ or $13/14$ floats.

For our and many other applications a large number of inversions are performed on a single gauge configuration. In this case, one can exploit the constant gauge field by grouping the random vectors in small bundles, thus applying the Dslash for multiple right-hand sides (rhs) at once:
\begin{equation}
    \left(w_x^{(1)}, w_x^{(2)}, \ldots, w_x^{(n)}\right) = D_{x,x'} \left(v_{x'}^{(1)}, v_{x'}^{(2)}, \ldots, v_x^{(n)}\right)\ .
\end{equation}
This increases the arithmetic intensity of the HISQ Dslash as the load of the gauge field occurs only once for the $n$ rhs. 
\begin{table}
    \centering
    \begin{tabular}{cccccccccc}\toprule
        \#rhs && 1 & 2 & 3 & 4 & 5 &6 & 8\\ \midrule
        $\mathrm{Flop/byte}$ (full)&& 0.73 & 1.16 & 1.45 & 1.65 & 1.80 & 1.91 & 2.08 \\ 
        $\mathrm{Flop/byte}$ (r14)&&0.80 &1.25 & 1.53 & 1.73 & 1.87 &1.98 & 2.14\\\bottomrule
    \end{tabular}
    \caption{\label{arithm_intens}The arithmetic intensity of the HISQ Dslash for different number of right-hand sides (rhs) using full or reduced 14 float storage (r14) for the Naik links.}
\end{table}
Increasing the number of rhs from 1 to 4 already results in an improvement by a factor of more than 2. In the limiting case of assuming the gauge fields do not have to be loaded at all, the highest arithmetic intensity that can be reached is $\sim\! 2.75$. At $n=8$ we have reached already $\sim\! 75\%$ of the limiting peak intensity, while for 1 rhs we only obtain $25\! -\! 30\%$. For an increasing number of rhs the memory transfers caused by loading the gauge fields are no longer dominating and thus also the impact of reconstructing the Naik links is less pronounced. 
Note that all numbers given here, as well as the performance data in the following, are for single-precision computations. However, the arguments work also for double-precision and the arithmetic intensity, in this case, is just half of the one in the single-precision case.
For the full inverter the linear algebra operations in the conjugate gradient do not allow for the reuse of any constant fields. They therefore limit the achievable speedup.
 
\begin{table}[h]
    \centering
    \begin{tabular}{lcccc} \toprule
                                           & 5110P     & 7120P     & K20       & K40         \\\midrule
        Cores / SMX                        & 60        & 61        & 13        & 15          \\
        (Threads/Core) / (Cores/SMX)       & 4         & 4         & 192       & 192         \\
        Clock Speed [$\mathrm{MHz}$]       & 1053      & 1238/1333 & 706       & 745/810/875 \\
        L1 Cache / Core [$\mathrm{KB}$]    & 32        & 32        & 16-48     & 16-48       \\
        L2 Cache [$\mathrm{MB}$]           & 30        & 30.5      & 1.5       & 1.5         \\
        Memory Size [$\mathrm{GB}$]        & 8         & 16        & 5         & 12          \\
        peak fp32/64 [$\mathrm{TFlop/s}$]  & 2.02/1.01 & 2.42/1.21 & 3.52/1.17 & 4.29/1.43   \\
        Memory Bandwidth [$\mathrm{GB/s}$] & 320       & 352       & 208       & 288         \\
\bottomrule
    \end{tabular}
    \caption{\label{tab2}The important technical data of the accelerators we have used in our benchmarks.}\
\vspace{-2em}
\end{table}

We summarized some technical data of the accelerators we use for our comparison in Table~\ref{tab2}. In the following we will only discuss our implementation for the Intel\SymbReg Xeon Phi\SymbTra. Information about our GPU implementation can be found in~\cite{gpu_mic_lat14}. 

\section{MIC}

\begin{figure}[h]
    \centering
    \includegraphics[width=0.87\textwidth]{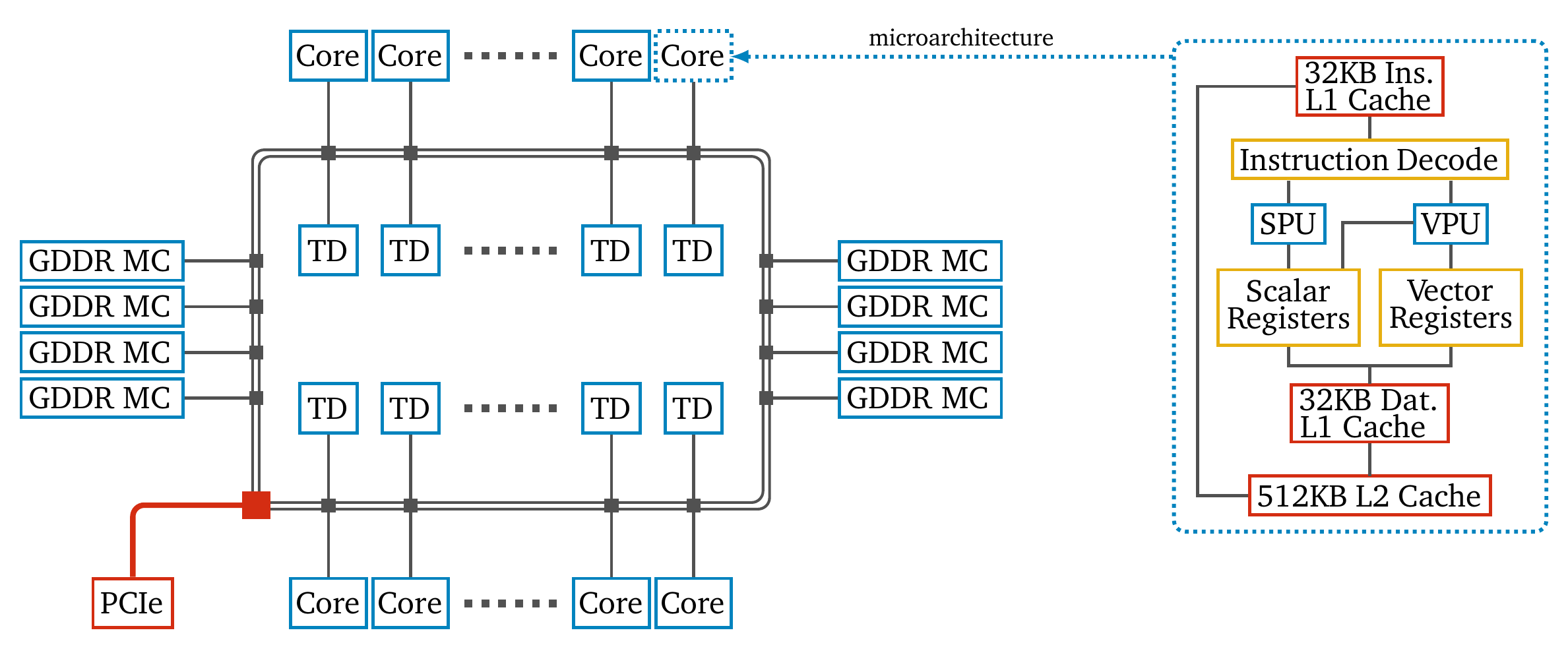}
    \caption{\label{mic_ring} Visualization of the bidirectional ring on the die and the microarchitecture of one core showing the Scalar Processing Unit (SPU), Vector Processing Unit (VPU) and the cache hierarchy. The latter is kept fully coherent through global distributed tag directories (TD).}
\end{figure}

The Intel\SymbReg Xeon Phi\SymbTra is an in-order \texttt{x}86 based many-core processor~\cite{mic}. The coprocessor runs a Linux $\mu$OS and can have up to 61 cores combined via a bidirectional ring (see Fig.~\ref{mic_ring}). Therefore, the memory transfers are limited by concurrency reaching only $149\;\mathrm{GB/s}$ on a 7120P running a stream triad benchmark~\cite{stream_mic}. Each core has a private L1 data and instruction cache and a globally visible L2 cache. In the case of an local L2 cache miss, a core can cross-snoop another's core L2 cache and if the data is present avoid a direct memory access.
Each core has thirty-two $512\;\mathrm{bit}$ \texttt{zmm} vector registers and 4 hardware context threads. To fully utilize the Many Integrated Core (MIC) it is, especially for memory-bound applications, necessary to run with four threads per core. This offers more flexibility to the processor to swap the context of a thread, which is currently stalled by a cache miss. The MIC has its own SIMD instruction set extension \texttt{IMIC} with support for fused multiply-add and masked instructions. The latter allows to conditionally execute vector instructions on single elements of a vector register. The coprocessor can stream data directly into memory without reading the original content of an entire cache line, thus bypassing the cache and increasing the performance of algorithms where the memory footprint is too large for the cache.

{\bf Implementation:} We have parallelized our program with OpenMP and vectorized it using low-level compiler functions  called intrinsics. These are expanded inline and do not require explicit register management or instruction scheduling through the programmer as in pure assembly code. There are $512\;\mathrm{bit}$ intrinsics data types for single- and double-precision accuracy as well as for integer values. More than 32 variables of a $512\;\mathrm{bit}$ data type can be used simultaneously. With only 32 \texttt{zmm} registers available in hardware, the compiler is, in this case, forced to use ``spills'', i.e.\ temporarily storing the contents of a register into L1 cache, and reloading the register when the data is required later, thereby increasing memory bandwidth pressure and cache pollution. When using intrinsics the software has to be designed in a register aware manner; only the explicit management of the registers is taken over by the compiler. We found that the compiler is only able to optimize code over small regions. Thus, the order of intrinsics can have an influence on the achieved performance, thereby making optimizations more difficult. Nonetheless, the use of intrinsics for the Dslash kernel is lightweight requiring only a subset of 9 instructions. Due to the different links needed for the nearest and third-nearest neighbor term we implemented both in separate kernels, thereby reducing cache pollution and simplifying cache reuse for the vectors. For the global sums inside the linear algebra kernels we use the OpenMP reduction clause. In order to avoid explicit barriers, each thread repeats the calculation of the coefficients necessary for the CG in a thread local variable.  

{\bf Site fusion:} One problem of using $512\;\mathrm{bit}$ registers involving $\mathrm{SU}(3)$ matrix-vector products is that one matrix/vector does not fit into an integer number of \texttt{zmm} registers without padding.
\begin{figure}[h]
    \centering
    \includegraphics[width=.77\textwidth]{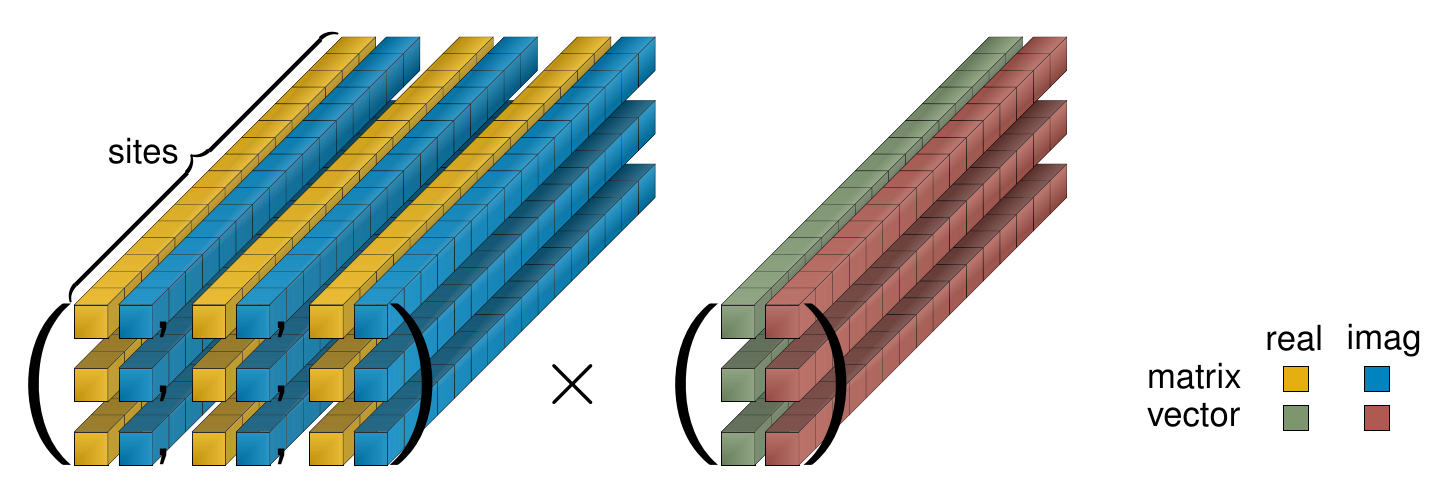}
    \caption{\label{16fold} Visualization of a fused matrix-vector product using the 16-fold vectorization scheme. Each lane corresponds to one \texttt{zmm} register, which holds the same element of all 16 sites.}
\end{figure}
Because of this, it is more efficient to process several matrix-vector products at the same time using a site fusion method. A naive single-precision implementation could be to create a ``Struct of Arrays'' (SoA) object for 16 matrices as well as for 16 vectors. Such a SoA vector object requires 6 \texttt{zmm} registers when it is loaded from memory. One specific register then refers to the real or imaginary part of the same color component gathered from all 16 vectors, thus each vector register can be treated in a ``scalar way'' (see Fig.~\ref{16fold}). These SoA objects are stored in an array using a site ordering technique. Our Dslash kernel runs best with streaming through $xy$-planes and is specifically adapted for inverting multiple right-hand sides. Therefore, we use a 8-fold site fusion method, combining 8 sites of the same parity in $x$-direction, which makes the matrix-vector products less trivial and requires explicit in-register align/blend operations. By doing so, we reduce the register pressure by 50\% compared to the naive 16-fold site fusion method, leaving more space for the intermediate result of the right-hand sides for each direction $\mu$. This is why the 8-fold site fusion is $55\%$ faster compared to the 16-fold scheme at 4 rhs (see Fig.~\ref{perf_changes}). For one right-hand side this optimization is insignificant since the 16-fold matrix-vector product requires only 30 of the 32 in hardware available \texttt{zmm} registers.
\begin{figure}[t]
    \centering
    \includegraphics[width=.46\textwidth]{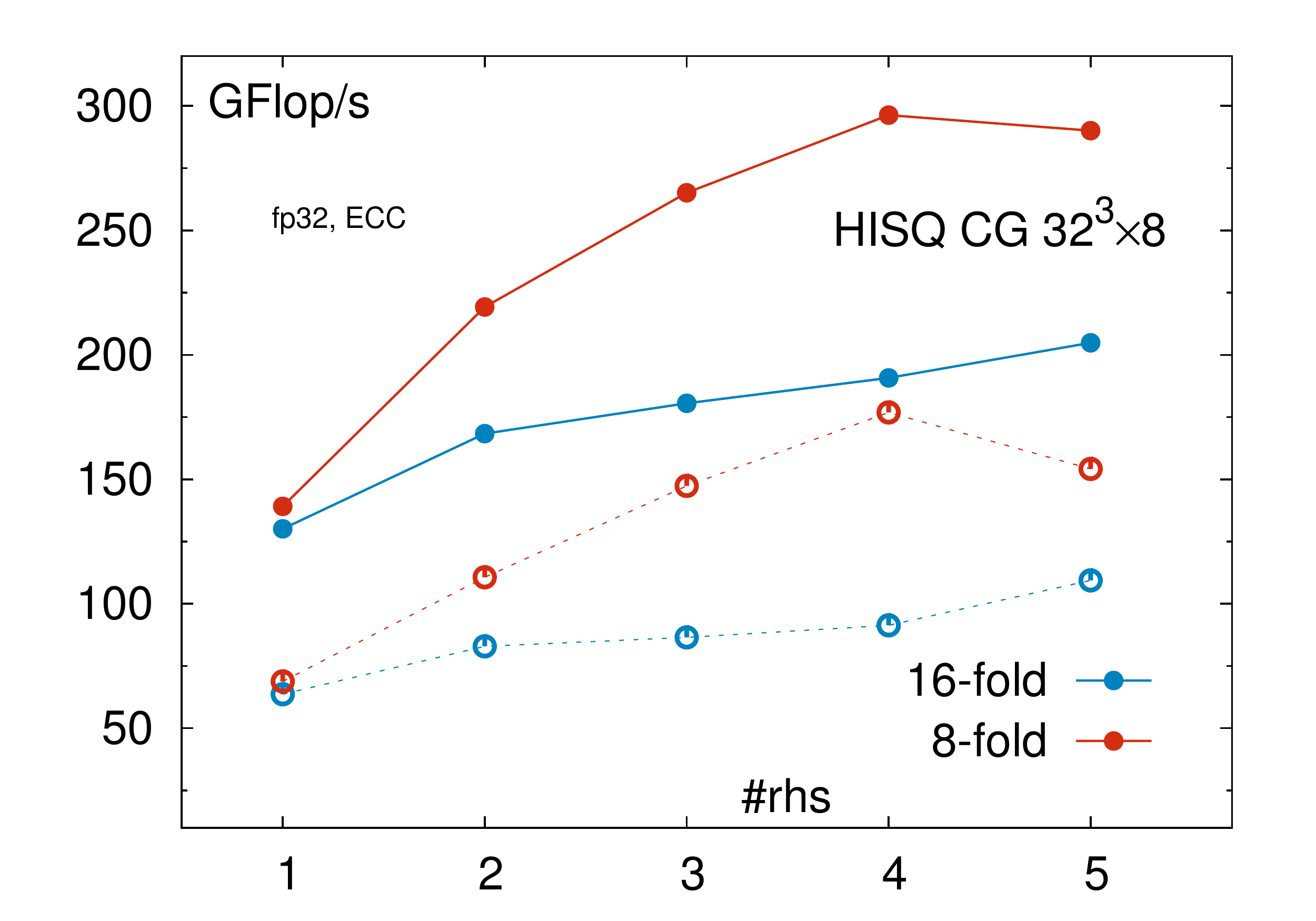}
    \caption{\label{perf_changes}Performance comparison of the 8-fold and 16-fold vectorization scheme measured on a 5110P with enabled ECC. The dashed lines correspond to kernels without software prefetching. }
\end{figure}

{\bf Prefetching:} For indirect memory access, i.e.\ the array index is a non-trivial calculation or loaded from memory, it is not possible for the compiler to insert software prefetches. The MIC has a L2 hardware prefetcher which is able to recognize simple access pattern. We found that it does a good job for a linear memory access. Thus, there is no need for software prefetching by hand inside the linear algebra operations of the CG. However, the access pattern of the Dslash kernel is too complicated for the hardware prefetcher. Therefore, it is required to insert software prefetches using intrinsics. The inverter runs $2\times$ faster with inserted software prefetches. We unroll the loop over all directions and prefetch always one $\mu$-term ahead. The first right-hand side vector and link of the first $\mu$-term are prefetched at the end of the previous site loop iteration. Considering that there is no reuse of gauge links, it is more efficient to prefetch these into the non-temporal cache. For the vectors we use temporal prefetch hints. It is important to note that software prefeteches are dropped if they cause a page table walk and in order to counterbalance the increased TLB pressure from the multiple right-hand sides, we store, for each lattice site, all rhs contiguously in memory. This approach is 15\% faster for large lattices compared to an implementation which stores each right-hand side in a separate array.

\section{Comparison}

For the Xeon Phi\SymbTra we used the Intel\SymbReg Compiler 14.0 and MPSS 3.3 with disabled instruction cache snooping, huge pages and a balanced processor affinity. For the GPU part we used the NVIDIA\SymbReg CUDA 6.0 toolkit. For the K40 we enabled GPU boost at the highest possible clock rate 875 MHz. In all benchmarks we left ECC enabled.

\begin{figure}[t]
    \centering
    \includegraphics[width=.496
    \textwidth]{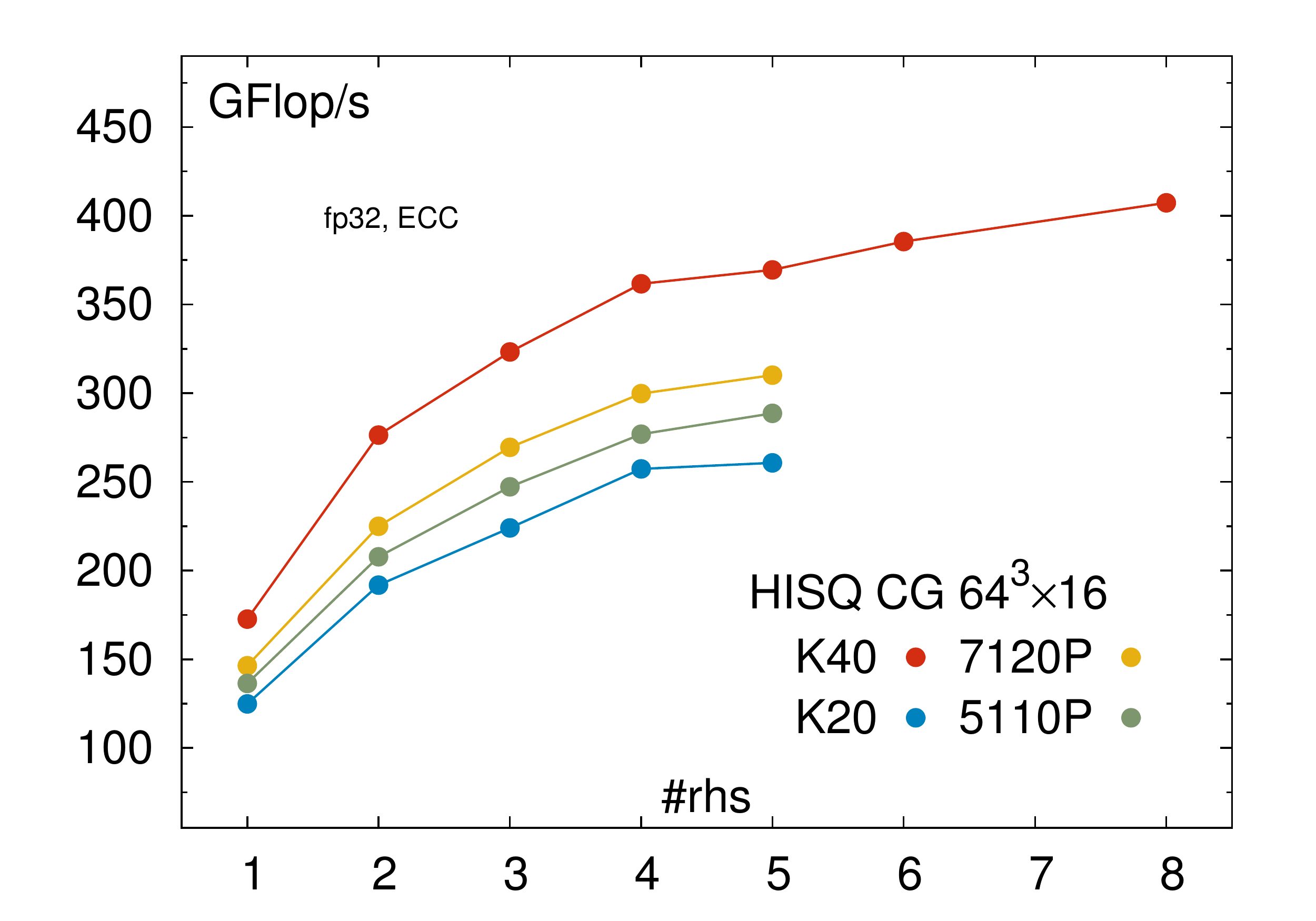}
    \includegraphics[width=.496\textwidth]{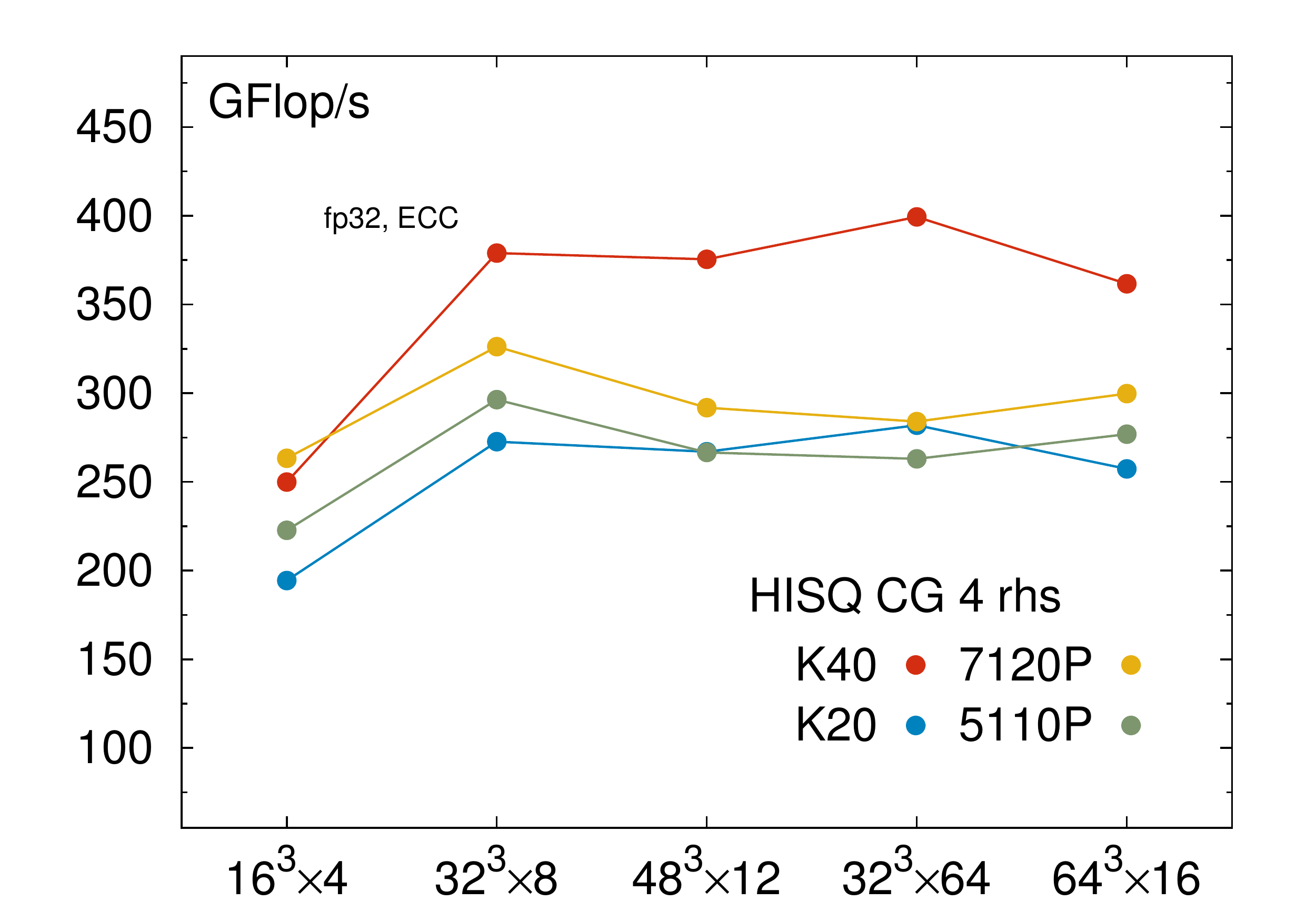}
    \caption{\label{fig4}Performance of the HISQ inverter on different accelerators for a $64^3\!\!\times\!\!16$ lattice as a function of the number of rhs (left) and for 4 rhs as a function of the lattice size (right). 
     }
\end{figure}

In the left panel of Fig.~\ref{fig4} we show the performance as a function of the number of rhs. The maximum number of rhs is  limited by memory requirements. We observe roughly the behavior as expected from the increased arithmetic intensity. Comparing the results using four right-hand sides to one right-hand side we find a speedup of roughly 2.05 for the full CG, very close to the increase of 2.16 in arithmetic intensity for the Dslash. Despite the linear algebra operations that limit the obtainable speedup this is still about 95\% of the expected effect.  Independent of the number of rhs the ordering with decreasing performance is K40, 7120P, 5110P, K20 with the relative performance  roughly given by 1.4, 1.2, 1.1, 1.0 (normalized to K20). 

In the right panel we show the performance with 4 rhs as a function of the lattice size. Ignoring the smallest size, the K40 is always superior while the 7120P is faster than the K20. The $32^3\!\!\times\!\!64$ lattice seems to be more expedient for the  SIMT model of the GPU.  

{\bf Acknowledgments:} We would like to thank Intel\SymbReg for providing a Xeon Phi\SymbTra system and especially Jeongnim Kim for supporting this work. We acknowledge support from NVIDIA\SymbReg through the CUDA Research Center program.

\end{document}